%% file: svmult/author/main_chapter.tex
\documentclass[graybox]{svmult/svmult}

\usepackage{type1cm}        
\usepackage{makeidx}         
\usepackage{graphicx}        
\usepackage{multicol}        
\usepackage[bottom]{footmisc}

\usepackage{newtxtext}       %
\usepackage[varvw]{newtxmath}       
\usepackage[numbers]{natbib}

\makeindex             

\begin{document}

\title*{Machine Learning as a Transformative Tool for (Exo-)Planetary Science}
\author{Jeanne Davoult\orcidID{0000-0002-6177-2085}, Valentin T. Bickel\orcidID{0000-0002-7914-2516}, Caroline Haslebacher\orcidID{0000-0003-0674-1216}, Yann Alibert\orcidID{0000-0002-4644-8818}, Daniel Angerhausen\orcidID{0000-0001-6138-8633}, Carles Cantero\orcidID{0000-0003-2073-782X}, Jo Ann Egger\orcidID{0000-0003-1628-4231}, Romain Eltschinger\orcidID{0009-0001-7351-3232}, Yannick Eyholzer\orcidID{0009-0004-3041-2249}, Emily O. Garvin\orcidID{0000-0003-2530-9330}, Salome Gruchola\orcidID{0000-0002-9757-1402}, Adrien Leleu\orcidID{0000-0003-2051-7974}, Sara Marques\orcidID{0009-0007-3825-7791}, and Yinan Zhao\orcidID{0000-0002-5009-4645}}

\authorrunning{Davoult et al.}

\institute{Jeanne Davoult \at Institut of Space Research, German Aerospace Center (DLR), Rutherfordstrasse 2, 12489 Berlin, Germany, \at Division of Space Research and Planetary Sciences, Physics Institute, University of Bern, Sidlerstrasse 5, 3012 Bern, Switzerland, \email{jeanne.davoult@dlr.de}
\and Valentin T. Bickel \at Center for Space and Habitability, University of Bern, Gesellschaftsstrasse 6, 3012 Bern, Switzerland, \email{valentin.bickel@unibe.ch}
\and Caroline Haslebacher \at Division of Space Research and Planetary Sciences, Physics Institute, University of Bern, Sidlerstrasse 5, 3012 Bern, Switzerland, \at and Southwest Research Institute, 1301 Walnut St Suite 400, 80302 Boulder, CO, USA, \email{caroline.haslebacher@contractor.swri.org}
\and Yann Alibert \at Center for Space and Habitability, University of Bern, Gesellschaftsstrasse 6, 3012 Bern, Switzerland, \email{yann.alibert@unibe.ch}
\and Daniel Angerhausen \at ETH Zurich, Institute for Particle Physics and Astrophysics
Wolfgang-Pauli-Str. 27, 8093 Zurich, Switzerland, \at Blue Marble Space Institute of Science, Seattle, WA, USA, SETI Institute, 189 N. Bernado Ave, Mountain View, CA 94043, USA, \email{dangerhau@phys.ethz.ch}
\and Carles Cantero \at Observatoire de Genève, Chemin Pegasi 51, 1290 Versoix, Switzerland, \email{carles.canteromitjans@unige.ch}
\and Jo Ann Egger \at Division of Space Research and Planetary Sciences, Physics Institute, University of Bern, Sidlerstrasse 5, 3012 Bern, Switzerland, \email{jo-ann.egger@unibe.ch}
\and Romain Eltschinger \at Center for Space and Habitability, University of Bern, Gesellschaftsstrasse 6, 3012 Bern, Switzerland, \at Division of Space Research and Planetary Sciences, Physics Institute, University of Bern, Gesellschaftsstrasse 6, 3012 Bern, Switzerland, \email{romain.eltschinger@unibe.ch}
\and Yannick Eyholzer \at Observatoire de Gen\`eve, Universit\'e de Gen\`eve, Chemin Pegasi, 51, 1290 Versoix, Switzerland, \email{Yannick.Eyholzer@unige.ch}
\and Emily O. Garvin \at Institute for Particle Physics and Astrophysics, ETH Zürich, Wolfang-Pauli-Strasse 27, 8093 Zürich, Switzerland, \email{egarvin@phys.ethz.ch}
\and Salome Gruchola \at Division of Space Research and Planetary Sciences, Physics Institute, University of Bern, Sidlerstrasse 5, 3012 Bern, Switzerland, \email{salome.gruchola@unibe.ch}
\and Adrien Leleu \at Observatoire de Gen\`eve, Universit\'e de Gen\`eve, Chemin Pegasi, 51, 1290 Versoix, Switzerland, \email{adrien.leleu@unige.ch}
\and Sara Marques \at Center for Space and Habitability, University of Bern, Gesellschaftsstrasse 6, 3012 Bern, Switzerland, \email{sara.marques@unibe.ch}
\and Yinan Zhao \at Department of Astronomy, University of Texas at Austin, 2515 Speedway, Austin, TX 78712, USA, \email{yinan.zhao@austin.utexas.edu}
}
%
%
\maketitle


\abstract{The exploration of planetary bodies in our Solar system and beyond relies on the processing and interpretation of large, spatio-temporally inconsistent, and heterogeneous datasets. Recent advances in machine learning (ML) provide unprecedented opportunities to address many fundamental challenges posed by these heterogeneous and hyper-dimensional datasets. This review chapter highlights innovative ML methodologies that were developed and used by NCCR PlanetS members to address three overarching challenges in (exo)planetary science. The first challenge is sequence modelling, which encompasses the intricate analysis of one-dimensional data such as time series of radial velocities and light curves, among other examples. Secondly, there is pattern recognition that involves studying correlations, leveraging convolutional neural networks for feature extraction, mapping and cross correlation among other examples., anomaly detection through variational autoencoders, and unsupervised clustering of mass spectrometric data. Lastly, there are generative models and emulation-based Bayesian analysis, which encompass the development of predictive models for planetary interior structure, employing Deep Neural Networks to understand planet formation mechanisms. These innovative ML methodologies herald a paradigm shift in the processing of data and numerical models that represent inherent challenges in planetary and exoplanetary science, paving the way for revolutionary discoveries and ideas in this field.}

\section{Introduction}
\label{sec:intro}

Planetary sciences have evolved over the centuries, transitioning from purely astronomical observations to a multidisciplinary field including physics, chemistry, geology, and computer science. The advent of space exploration marked a major turning point, with missions such as Mariner \citep[1960s-70s; ][]{Abelson1965,Dunne1978,Bailey2013} opening the path to the exploration of Mars, Venus, and Mercury, followed by Voyager 1 and 2 \citep[1977; ][]{Abelson1977,Kohlhase1977}, which provided unprecedented insights into the outer planets. More recently, exoplanet-focused missions like Kepler \citep[2009; ][]{Borucki2010} and the Transiting Exoplanet Survey Satellite \citep[TESS, 2018; ][]{Ricker2015} have largely expanded our catalog of known worlds with thousands of new detections, while missions such as Juno \citep[2011; ][]{Bolton2017} and the James Webb Space Telescope \citep[JWST, 2021; ][]{Gardner2006} have deepened our understanding of planetary and exoplanetary environments, and formation processes. In parallel, asteroid exploration missions like Origins Spectral Interpretation Resource Identification Security – Regolith Explorer \citep[OSIRIS-REx, 2016; ][]{Lauretta2017} and Double Asteroid Redirection Test \citep[DART, 2021; ][]{Cheng2018} have provided invaluable data on small bodies.\\
Alongside observational breakthroughs, theoretical models have also advanced significantly, refining our understanding of planetary formation and evolution. Models such as the Grand Tack \citep{Walsh2012} and the Nice Model \citep{Gomes2005,Tsiganis2005,Morbidelli2005} have reshaped our comprehension of the Solar System’s dynamical history, while 3-D atmospheric simulations using general circulation models \citep[GCMs; ][]{Adcroft2004,Richardson2007,Gilli2017} provide detailed insights into planetary climates. Currently, exoplanetary studies rely on time-consuming computational techniques such as Smoothed Particle Hydrodynamics (SPH) for simulating giant impacts for example \citep[e.g., ][]{Timpe2023,Meier2025} or Magneto-Hydrodynamics (MHD) simulations for modelling protoplanetary disks \citep[e.g., ][]{Weder2023}.\\
Thus, planetary sciences now face two major challenges: the exponential growth of data and the increasing complexity of models and datasets. Missions like the Lunar and Mars Reconnaissance Orbiters \citep[MRO, 2005; e.g.,][]{McEwen2007}, Gaia \citep{Gaia2016}, JWST and in the near future Europa Clipper \citep[2024;][]{Howell2020} and JUpiter ICy moons Explorer \citep[JUICE, 2023;][]{Grasset2013}, the PLAnetary Transits and Oscillations of stars mission \citep[PLATO; ][]{Rauer2014,Rauer2016} or the Atmospheric Remote-sensing Infrared Exoplanet Large-survey mission \citep[ARIEL; ][]{Tinetti2022}, generate or will generate vast amounts of data, making manual analysis increasingly unfeasible. Meanwhile, the computational cost of sophisticated physical models continues to rise, limiting their efficiency.\\
Machine Learning (ML) and, more specifically, Deep Learning (DL) provide powerful solutions to these challenges. ML enables computers to learn from data and make decisions, in a similar way to how humans think. Unlike regular astrophysics programming that follows set physical rules, ML finds patterns, makes predictions, and gets better by processing data. It works in two main ways: classification and regression. Classification groups data into distinct classes, corresponding to discrete values, while regression seeks to approximate a function linking continuous variables, in order to highlight the underlying mechanism. By automating data analysis, accelerating simulations, and detecting patterns that traditional methods might overlook, the integration of AI-driven methodologies into planetary sciences offer new opportunities to interpret complex datasets and improve predictive modeling. In this chapter, we define machine learning (ML) as models that do not rely on neural network architectures, for example ensemble methods such as random forests or boosting, clustering algorithms, or support vector machines (SVM). These models are usually less complex and often depend on manual data preparation. Deep learning (DL), by contrast, refers to models based on neural networks, which are typically more complex and can automatically extract structure from data. It is important to specify that these definitions are used to name these two concepts in this chapter, but that in general DL is a sub-ensemble of ML. Finally, ML and DL are distinguished between "supervised2 methods, where models train on labelled data in order to solve classification or regression tasks, and "unsupervised" methods, where models learn to identify structures or groupings in unlabelled data, as in the case of clustering or dimension reduction. \\
The use of machine learning in space science began in the 1990s but grew significantly in the 2000s with increased computing power and large astronomical surveys. One of the first key applications was the use of neural networks to classify galaxies from Sloan Digital Sky Survey \citep[SDSS, 2000][]{York2000} images \citep{Gulati2000,LiLu2025}. In planetary sciences, the first machine learning models were adopted in early 2000s to map the martian surface \citep{Bue2006,Stepinski2006,Stepinski2007,Stepinski2009} through support vector machines (SVMs), supervised and unsupervised methods and to automate Mars rovers in the recognition of minerals in spectra \citep{Gilmore2000,Bornstein2005,Gilmore2008} with neural networks (NNs). Since then, models such as random forests, SVMs, and especially deep neural networks (DNNs) have transformed planetary science data analysis. Today, these methods are more and more used for exoplanet detection and characterization, studying planetary surfaces, or modeling planetary atmospheres. In particular, deep learning, with convolutional neural networks (CNNs) and variational autoencoders (VAEs), has become essential for processing the vast amounts of data from space missions.\\
The NCCR PlanetS has played a significant role in encouraging the adoption of ML and DL in planetary sciences. By leveraging interdisciplinary expertise, PlanetS has facilitated major advances in exoplanet detection and characterization, planetary system dynamics, atmospheric modeling, and the study of small bodies such as asteroids and comets. Since its start in 2014---coinciding more or less with the rise of ML in astronomy---PlanetS has contributed to the democratization of ML and DL techniques, promoting their application in astrophysical data analysis, planetary structure modeling, and beyond.\\
This chapter provides a review of the main contributions of PlanetS members and associates in the application of ML and DL in planetary sciences, illustrating how these techniques are revolutionizing the field and shaping the future of planetary research. In Section \ref{sec:SeqMod}, we explore research on sequence modeling, followed by an examination of pattern recognition in Section \ref{sec:PatRec}. Section \ref{sec:GenMod} introduces two studies that facilitate the emulation of internal structural models. Finally, we discuss the new challenges and limitations that the use of ML and DL in planetary science brings and conclude in Section \ref{sec:discussion}.



\section{Sequence modelling}
\label{sec:SeqMod}
In planetary science, sequence modeling is used to analyze and predict temporal, spatial, spectral or event-based sequences. Temporal sequence analysis includes studying radial velocity and stellar light curves to detect and characterize exoplanets, model observational noise, and track planetary atmospheric changes due to seasonal and weather variations. Spatial sequence analysis examines temperature and pressure profiles in atmospheres and the internal structure of planets, exoplanets, and moons. Spectral sequence analysis involves exoplanetary atmospheric absorption spectroscopy and the mineralogical composition of planetary surfaces. Event sequence analysis includes volcanic eruptions on Io per instance, or the frequency of asteroid impacts on the Moon. In addition, sequence modelling can be used to emulate results of theoretical models (surrogate models), as long as their predictions can be represented as sequences of any kind. An example developed in the NCCR PlanetS is presented at the end of this section.\\
Members of NCCR PlanetS have contributed to incorporate ML techniques in sequence modeling studies in various ways, as described in the following sections: noise modeling in stellar radial velocity time series (Sect. \ref{subsec:RVtimeseries}), transit timing variation (TTV) detection (Sect. \ref{subsec:TTV}), development of surrogate model for global models of planetary system formation (Sect. \ref{subsec:Transformer}) and detection of exoplanets in high-contrast imaging using angular differential imaging (Sect. \ref{subsec:hc-imaging}).

\subsection{Improving Earth-like planet detection in radial velocity using deep learning}
\label{subsec:RVtimeseries}
\input{svmult/author/Yinan/Yinan}

\subsection{Alleviating the transit timing variation bias in transit surveys}
\label{subsec:TTV}
\input{svmult/author/Yannick/Yannick}

\subsection{Surrogate models: using transformers to predict planetary systems architectures}
\label{subsec:Transformer}
\input{svmult/author/Yann/transformer}
\subsection{Machine learning for exoplanet detection in high-contrast imaging with angular differential imaging strategy}\label{subsec:hc-imaging}
\input{svmult/author/Carles/carles}

\section{Pattern recognition}
Pattern recognition has been one of the earliest applications of ML in planetary science, in early 2000s. Pattern recognition is the process of identifying and classifying regularities or structures in data, such as shapes, sequences, or trends. It involves training models to automatically detect these patterns and make predictions or decisions based on new, unseen data. It was first used for surface mapping, particularly on Mars, where ML techniques helped automate the classification of terrains or the recognition of craters using orbital imagery from missions like Mars Global Surveyor or Mars Express \citep{Bue2006,Stepinski2006,Stepinski2007,Stepinski2009}, or helped the identification of minerals in spectra \citep{Gilmore2000, Bornstein2005,Gilmore2008}. Since then, pattern recognition has been widely applied across different domains. Techniques such as neural networks, SVMs, clustering algorithms or ensemble learning methods have enabled the analyze of large datasets efficiently, uncovering patterns that would be difficult to identify manually.\\
PlanetS members have contributed to the use of ML techniques in pattern recognition with different topics: unveiling correlations between planets' properties in planetary systems (Sect. \ref{subsec:correlations}), characterizing and mapping planetary surfaces (Sect. \ref{subsec:mapping}), detection of exoplanets in high-contrast spectroscopy (Sect. \ref{subsec:hc-spectroscopy}), and analyses of mass spectra (Sect. \ref{subsec:massspectra}).\\

\label{sec:PatRec}
\subsection{Unveiling correlations in planetary systems}\label{subsec:correlations}
\input{svmult/author/Jeanne/jeanne}

\subsection{Features extraction}\label{subsec:mapping}
\subsubsection{Characterizing the spatio-temporal evolution of planetary surfaces with Deep Learning}
\input{svmult/author/Valentin/main_valentin}
\subsubsection{Mapping linear surface features on planetary bodies with deep learning}
\input{svmult/author/Caroline/main_caroline} 
\subsection{Machine learning for exoplanet detection in high-contrast spectroscopy} \label{subsec:hc-spectroscopy}
\input{svmult/author/Emily/main_emily}
\subsection{Unsupervised techniques for \textit{in situ} mass spectrometry}
\label{subsec:massspectra}
\input{svmult/author/Salome/main_salome}

\section{Model emulation with deep learning} \label{sec:GenMod}
In recent years, the continuous improvement of numerical models in planetary sciences has led to increasingly complex and computationally expensive simulations. These models, while more accurate, often require significant time and resources to run. To address this challenge, deep learning techniques are being employed to learn from the outputs of these time-consuming models and predict results without the need to rerun the full numerical model. This approach, often referred to as surrogate modeling or emulation, is becoming more common, particularly in domains where large parameter exploration or real-time predictions are needed. Using neural network to approximate the behavior of complex models can drastically reduce computation times, allowing the growth of numerical simulations.\\
PlanetS members have used model emulation to predict architecture of planetary systems as seen in Sect. \ref{subsec:Transformer}, but also to predict planetary radii from internal structure model (Sect. \ref{subsec:planetic}) and planetary mass from planetary formation model (Sect. \ref{subsec:DNN_mass}).\\


\subsection{plaNETic: a predictive model for planetary interior structure} \label{subsec:planetic}
\input{svmult/author/JoAnn/main_joann}

\subsection{Accelerating planetary system formation models with DNNs} \label{subsec:DNN_mass}
\input{svmult/author/Yann/mass_prediction}

\section{Discussion and conclusion}\label{sec:discussion}
Machine Learning is fundamentally changing scientific research, including (exo)plane\-tary sciences, by enabling the processing and analysis of vast datasets at unprecedented speeds and scales. Machine learning helps automate tedious, time-consuming tasks, like the identification of small geologic features in planetary image datasets, and significantly lowers the computing costs of complex models. Over the past few decades, ML has changed how we analyze data by making it easier to handle complex, heterogeneous, multi- and hyper-dimensional information and recognize patterns that might be, or a highly likely to be, missed by humans. More specifically, ML provides new approaches for analyzing atmospheric compositions \citep[e.g., ][]{Giobergia2023,Forestano2023,DuqueCastano2025,Dahlbudding2024,Gebhard2025} or helping prioritize targets for follow-up observations \citep[e.g., ][]{DuqueCastano2025, Davoult2025} 
Compared to traditional statistical methods, ML models show generally stronger predictive capabilities and can handle larger, more heterogeneous, and more complex datasets. On the other hand, traditional statistical methods provide clear insights and straightforward explainability about what the model has learned and offer better interpretability and theoretical understanding, while ML models can behave like `black boxes'. This underlines how the use of data-driven approaches has to remain a trade-off between model complexity and scientific interpretability.\\
Many challenges in using ML for (exo)planetary science remain, especially due to the limited amount and quality of available data. In many areas of (exo)planetary science, there are not enough data for training models---whether it is the number of known exoplanets, labeled images of planetary surfaces, asteroid observations, or atmospheric spectra. On top of that, data often come with noise, vary in quality, and suffer from other instrument-related limitations. Because of this, scientists increasingly rely on synthetic data to fill the gaps, but that might lead to training on data that are not exactly like reality and a difficulty to verify hypothesis due to lack of actual ground-truth data \citep[see e.g.,][]{Davoult2025}. Finally, models also risk overfitting, meaning they work well on the training data but fail when applied to new situations or larger datasets.\\
To move forward, we need ML models that are easier to interpret, and more and better data, which may come from upcoming space missions like PLATO, ARIEL, Europa Clipper, and BepiColombo \citep{Bepicolombo}, or new ground-based telescopes like the ELT\\
As ML changes (exo)planetary science, it also raises ethical issues. We need transparency in how ML methods and results are shared, and human insight and intuition should remain central in scientific work. There are also risks, like hidden biases in training data or incorrect results that could mislead research. It is also important to mention that most ML and DL algorithms currently offer limited mechanisms for incorporating uncertainties or propagating errors. This limitation is particularly relevant in scientific contexts, where quantifying uncertainty is often as critical as producing accurate predictions. Bridging this gap remains an important work for future ML and DL research and development.\\
Even with these challenges, ML and DL are reshaping how we explore space and plan missions. New algorithms help create better target lists for missions like JWST, the Habitable Worlds Observatory \citep[HWO; ][]{Harada2024,Stark2024}, and the Trace Gas Orbiter \citep[TGO, 2016; ][]{TGO}, making better use of valuable telescope time \citep[e.g., ][]{garvin2024machine,Torres-Quijano2025, Davoult2025}, supporting real-time decisions during missions \citep[recently e.g., ][]{Gruchola2024, McDonnell2024,Cao2025}, and improving how we design observations \citep[e.g., ][]{Li2025}.\\
This chapter presents twelve different areas in which PlanetS members have contributed by incorporating ML models to their work. It encompasses projects using sequence modeling such as the modeling of stellar activity in RV time-series, the detection of TTVs in stellar light-curves, the detection of exoplanets in high-contrast imaging using ADI, and the modeling of planetary systems as a 2D-sequence. We also gathered projects using pattern recognition such as unveiling correlations in planetary systems, mapping planetary surfaces, the search for patterns linked to molecular signatures in cross correlation in high-contrast spectroscopy, and clustering techniques for mass spectra. Finally, we present works using neural networks for model emulation to predict planetary radius from internal structure models and planetary mass from planetary formation models.\\
Despite new challenges arising with the increasing use of ML and DL models, those techniques can help reach goals precedently unreachable. For example, the improvement of Earth-like planet detection \citep[e.g., ][]{Armstrong2021,Zhao2024,Roche2024,Hansen2017, Torres-Quijano2025} can ultimately allow us to detect an exoplanet favourable to host life. Another example is the scanning of complete datasets of planetary surfaces orders of magnitude faster than a human.\\

Machine learning is changing the way we analyze planetary data and expanding the kinds of questions we can ask about the origins and future of the solar system and others. As our tools become more advanced, so does our ability to explore and understand planetary bodies.

\bibliographystyle{svmult/author/spphys_ed}
\bibliography{ref.bib}
\end{document}

%% file: svmult/author/Yinan/Yinan.tex
\begin{figure}[b]
\sidecaption
\includegraphics[width=13cm]{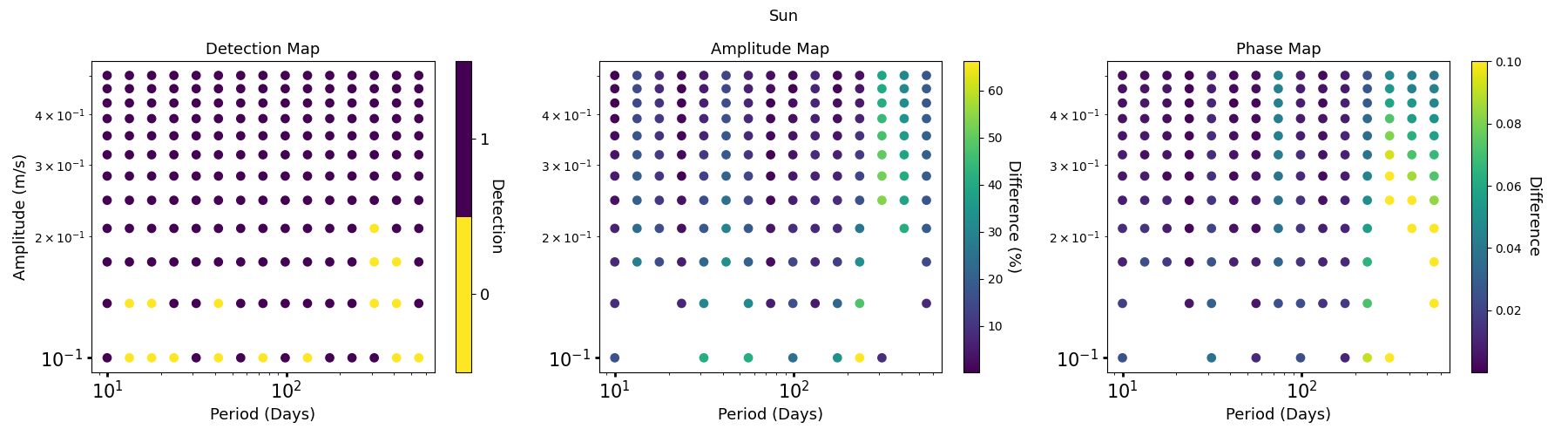}
\caption{Exoplanet detection limits from the neural network framework in the case of HARPS-N solar spectra. 
Left: Detection limit map in the period-amplitude domain. The red dots indicate the successful detection by the trained NN with a false alarm probability (FAP) $>$ 0.1\%. Middle: Amplitude comparison between the injected and recovered signals in the period-amplitude domain. The amplitude difference for most of the recovered signals are $<$ 20\%. The large difference at low amplitude (0.1 m/s) is likely due to noise in the data. Right: Phase comparison between the injected and recovered signals in the period-amplitude domain. The phase difference for most of the recovered planets is $<$ 0.04. The large differences at long periods is likely due to a poorer sampling of the phase for long-period planetary signals. Figure adapted from \cite{Zhao2024}.}
\label{fig:Zhao2}
\end{figure}

The detection of low-mass planets using the radial velocity (RV) technique remains extremely challenging in the presence of stellar activity. Many novel methods have been proposed to mitigate stellar activity for exoplanet detection as the presence of stellar activity in RV measurements is the current major limitation. Both the template matching algorithm and the cross-correlation function (CCF) algorithm extract the average RV information contained in a spectrum. Key variations at the spectral level related to stellar activity may be lost when performing the dimensionality reduction imposed by the RV extraction algorithms. In \cite{Zhao2024}, the authors present a novel convolutional neural network (CNN)-based algorithm that efficiently models stellar activity signals at the spectral level, enhancing the detection of Earth-like planets.\\
A CNN is trained to build the correlation between the spectral line profile change and the corresponding RV, full width at half maximum (FWHM) and bisector span (BIS) values derived from the classical cross-correlation function. The RV signals only due to the doppler effect can be obtained by subtracting the RV time series predicted from the trained NN (due to stellar activity) from the raw RV time series. The residuals is mainly due to the presence of planets.\\
In the context of this work, the main challenge lies in the amount of data that can be used to train CNNs for different types of stars. The number of spectra available for most of RV observations of different stars is less than 1000. Hence the limitation of the available data for different stars is therefore the main obstacle. In order to efficiently train the NN, the authors use cross-validation as a resampling method. In detail, the spectral shells are randomly divided into 10 groups. For each iteration, one group is designated as the test set, while the remaining groups are used to train a model. The final results is the summary of the evaluations conducted on the test sets across all iterations.\\
In \cite{Zhao2024}, the authors aim to build a NN that can predict the RV induced by stellar activity from spectral shape shells in the $\left(\frac{df_0}{dv},f_0\right)$ space. Once the NN is well trained and insensitive to planetary signals, then simply removing the predicted RVs from the raw RVs should enhance the significance of planet signals embedded into stellar signals. 
This algorithm has been tested on three intensively observed stars: Alpha Centauri B (HD 128621), Tau ceti (HD 10700), and the Sun. By injecting simulated planetary signals at the spectral level, the authors of \cite{Zhao2024} demonstrate that their machine learning algorithm can achieve, for HD 128621 and HD 10700, a detection threshold of 0.5 m/s in semi-amplitude for planets with periods ranging from 10 to 300 days. This threshold would correspond to the detection of a $\sim$ 4~$M_\oplus$ in the habitable zone of those stars. As an example, the framework used ~380 spectra for HD 10700. On the HARPS-N solar dataset, thanks to significantly more data, their algorithm is even more efficient at mitigating stellar activity signals and can reach a threshold of 0.2 m/s, which would correspond to a 2.2~$M_\oplus$ planet on the orbit of the Earth as shown in Fig. \ref{fig:Zhao2}. To the best of our knowledge, it is the first time that such low detection thresholds are reported for the Sun, but also for other stars, and therefore this highlights the efficiency of this CNN-based algorithm at mitigating stellar activity in RV measurements.


%% file: svmult/author/Yannick/Yannick.tex
The transit method has proven to be the most successful method for the discovery of exoplanets so far. By continuously monitoring the stellar flux over an extended duration, it is possible to detect periodic dips in the star brightness, indicative of a planet transiting across the star along the observer's line of sight. These dips of flux allow the measurement of the relative size of the planet to the host star. When a single planet orbits a single star, its orbit is periodic, which implies that the transit happens at a fixed time interval.\\
However, in multi-planetary systems the assumption of strictly Keplerian orbits no longer holds, as mutual gravitational interactions between the planets induce perturbations \citep{agol2005}. The magnitude of these perturbations depends on the system's configuration, and is significantly amplified for (near)-resonant systems as the perturbative effects accumulate over time \citep{Lithwick_2012,Nesvorný_2014,Agol_2016}. This often leads to quasi-periodic orbits, deviating from the Keplerian periodicity. Observationally, these deviations manifest as transit timing variations (TTVs). Standard algorithms rely on the periodicity of the signal and typically fail in this situation and suffer strong biases. This is unfortunate as TTVs offer a wealth of information; they encode gravitational interractions, allowing the retrieve the masses of the planets \citep[see e.g.; ][]{Nesvorny2013},and they allow to infer the presence of a non-transiting planet \citep{Xie_2014,Zhu_2018}. More importantly, systems in resonance often kept a pristine configuration after their formation, thus detecting them offer a way to explore system formation histories. \\ 
In this context, RIVERS.deep \citep{Leleu2021b} is an approach within the broader RIVERS framework (Eyholzer et al. in prep) that aims to tackle the challenge of TTVs. Specifically, RIVERS.deep is designed to identify planetary transit signals in 2D representation of light curves, known as river diagrams or river plots \citep{Carter2012-zv}, using computer vision through machine learning.\\
The RIVERS.deep model is a fully convolutional neural network based on the "Tiramisu" architecture \citep{Jegou2016}, a denser and more complex variant of the classical "U-net" \citep{Ronneberger2015}. 
The model is trained to recognize the vertical tracks within river diagrams and produces a mask highlighting the transit signal from the background noise. This enables the identification of individual transit times without assuming strict periodicity.\\ 
RIVERS.deep has enabled the detection of exoplanets missed by traditional methods, such as Kepler-1705 b and c \citep{Leleu2021b}, and Kepler-1972 b and c \citep{Leleu2022}---both resonant pairs affected by TTVs. It also improved the characterization of known systems, like Kepler-404, revealing resonant dynamics and clarifying a false positive as part of a TTV-affected signal (Eyholzer et al. in prep).Additionally, RIVERS.deep refined mass estimates for 34 sub-Neptunes \citep{Leleu2023}, showing that resonant planets tend to be less dense, highlighting the importance of accurate TTV modeling \citep{Leleu2024}.

%% file: svmult/author/Yann/transformer.tex
Sophisticated planetary system formation models have been developed in the NCCR PlanetS (see chapter Kessler et al. on the Bern model) in order to compute, in an end-to-end approach, the result of planet formation. Such a model essentially solves differential equations to predict, from the properties of a protoplanetary disk, populations of planetary systems in which all planetary properties (e.g. mass, orbital elements, composition, etc...) are computed in a self-consistent way. These results can then be used to interpret observations \citep[e.g. by predicting which planets are likely to exist in a system where only a few planets have been discovered, ][]{Davoult2024, Davoult2025}, and to constrain physical processes at work during planet formation \citep[e.g. by comparing model results with observations, ][]{Emsenhuber2025}. As multi-planetary systems contain, by definition, more than one planet, one key ingredient in the Bern model is the inclusion of a N-body code to compute the interactions between planets forming in the same system \citep{Alibert_etal_2013}. Combined with the fact that planets in a given system can have vastly different orbital period (which constrains the integration timescale to be small), the computation of a population of planetary systems requires massive computer resources. As an example, the NGPPS main population \citep[considering only 1000 systems with 100 forming planets each, ][]{Emsenhuber_etal_2021} took one million CPU hours to compute.\\ 
On the other hand, a planetary system can be considered either as a \emph{set} of objects, or as a \emph{sequence} of objects (provided an ordering scheme---e.g. based on the orbital period of planets---is fixed).
Such a sequence prediction model can, in turn, be used to emulate the results of the Bern model, thus accelerating largely the computation of planetary system architectures.\\
In a recent paper, Alibert, Davoult and Marques (2025, in review) have used this approach to train a transformer-based model that can be used to emulate some results of the Bern model. In this paper, the authors only considered the formation of planetary systems orbiting solar-type stars, and trained their model on a set of $\sim$25000 planetary systems computed with the Bern model, with 20 forming planets per system. In this paper, planets were characterized only by two features: their mass and semi-major axis.
The model is based on the transformer architecture \citep{Vaswani2017} that has become extremely popular thanks to the development of large language models (LLMs). In short, a transformer uses the so-called attention mechanism in order to model the probability distribution of the n-th element of a sequence, knowing the previous elements in the same sequence. Using terms appropriate for planetary systems, this means computing the probability distribution of the mass and semi-major axis a planet $n$ in a sequence (the planets being ordered by increasing semi-major axis), knowing the mass and semi-major axis of all planets $0$ to $n-1$ in the same system. 
Such a model, once trained, can be used to produce large number of sequences. All new sequences are always produced in an inside-out process, starting with the innermost planet and extending until the outer parts of the system. These sequence can then be used, for example, to predict what are the other planets that could reside, according to the Bern model, in the same system as some already discovered planets. 

%% file: svmult/author/Carles/carles.tex
The direct imaging of exoplanets using 10-meter-class ground-based telescopes has become a reality in modern astrophysics due to advancements in high-contrast imaging (HCI). Techniques such as extreme adaptive optics (AO) are used to correct atmospheric distortions, and state-of-the-art HCI instruments attached to the telescope employ coronagraphs \citep{Soummer2005Lyot} to block starlight and manage the high brightness contrast between stars and their exoplanets.\\
Despite these advanced technologies, exoplanet detection in high-contrast images remains challenging due to various noise sources. These include photon noise from residual stellar light and thermal background emission, as well as speckle noise arising from atmospheric turbulence and optical aberrations. Quasi-static speckles, which appear as scattered starlight artifacts, can closely resemble exoplanet signals in both shape and brightness, making it difficult to reliably distinguish true planetary signals from noise.\\
To overcome this, observing strategies like angular differential imaging \citep[ADI; ][]{marois2006angular} combined with powerful image post-processing algorithms are employed. ADI involves capturing a sequence of images in pupil-stabilized mode, where the telescope's pupil remains fixed while the image field rotates due to Earth’s motion. As a result, astrophysical signals rotate relative to quasi-static speckles, enabling to differentiate true exoplanets.

In \citep{Cantero2023}, the authors present NA-SODINN, a supervised deep learning classifier based on convolutional neural networks, specifically designed for robust exoplanet detection in ADI-PCA (PCA stands for principal component analysis) processed frames. 
Extensive evaluation, including performance assessments across diverse datasets from multiple instruments and participation in the Exoplanet Imaging Data Challenge \citep[EIDC][]{Cantalloube2020}, demonstrated that NA-SODINN consistently outperforms the most advanced detection algorithms. In Cantero et al. (in prep,a), the authors apply the NA-SODINN detector to reprocess the F150 sample from the SPHERE Infrared survey for Exoplanets \citep[SHINE; ][]{Desidera2021}. This sample comprises 150 targets observed with VLT/SPHERE using H2/H3 narrowband filters and angular differential imaging. By leveraging the enhanced detection capabilities of NA-SODINN, they aim to identify new planetary candidates within this dataset, advancing the search for exoplanets. The NA-SODINN detector improves detection limits in post-processed HCI data, allowing for the identification of a larger number of planetary candidates. This contribution is particularly important for filling the observational gap between RV and HCI surveys, providing crucial data for understanding exoplanet populations across a broader range of orbital separations.\\
In another paper, Cantero et al. (in prep, b), the authors present a novel speckle subtraction algorithm. Over the past decade, PCA has become the industry standard for speckle subtraction in HCI and is now widely integrated into most HCI software packages, including VIP \citep{Christiaens2023VIP}. Its widespread adoption is largely due to its simplicity and computational efficiency. However, PCA has two notable limitations: the loss of the planet signal \citep{Bonse2025} and the inability to account for non-linear correlations.\\
Cantero et al. (in prep., b) present an algorithm that addresses these challenges while achieving higher contrast levels. To this end, the authors are designing a deep learning-based denoiser model that learns the speckle noise distribution and significantly enhances subtraction performance. This approach combines several state-of-the-art deep learning technologies, including variational autoencoders and latent diffusion models, and leverages conditional learning for optimal results. The proposed model will be optimized for SPHERE and SPHERE+ data, as the next generation of instruments promises to significantly extend the reach of HCI observations.\\

%% file: svmult/author/Jeanne/jeanne.tex
The study of correlations between data is essential in science, as it enables trends to be identified and complex phenomena to be better understood. In the science of exoplanets, these analyses are crucial for gaining a better understanding of the formation and evolution of planetary systems, as well as refining the search for potentially habitable worlds. 
In recent years, several correlations have been identified among the growing population of exoplanets discovered by the CoRoT \citep{Auvergne2009}, Kepler, and more recently, TESS missions. One of the most well-established correlations is the increase in the frequency of giant planets with stellar metallicity \citep[e.g.,][]{Gonzalez1997, Santos2001}. This relationship is explained by the greater availability of heavy elements, which enhances accretion and facilitates the rapid formation of bodies reaching the critical mass \citep[e.g.,][]{Pollack1996}. Another correlation links the metallicity of the host star to the eccentricity of giant planets \citep[e.g.,][]{Dawson2013}. Additionally, the frequency of giant planets depends not only on stellar metallicity but also on stellar mass: it increases with stellar mass, which is assumed to be proportional to that of the protoplanetary disk \citep[e.g.,][]{Johnson2010, Bonfils2013} (see chapter Kessler et al., this collection). It has also been shown that stars hosting exoplanets exhibit a C/O abundance ratio that is 0.01 to 0.05 higher than that of stars without planets \citep{Melo2024} and tend to be refractory-depleted \citep{Yun2024}. 
Thus, it is clear that the properties of exoplanets are closely linked to the environment in which they form, including the host star, the protoplanetary disk, and planetary interactions. Studies based on synthetic populations of planetary systems generated by the Bern model \citep{Mishra2023b, Emsenhuber2023} have successfully connected the final architecture of exoplanetary systems to their formation pathways (see chapter Kessler et al. in prep on the Bern model). Furthermore, \cite{Davoult2024} identified correlations between the presence of an undetected Earth-like planet (ELP) in a system and various parameters, including the host star’s mass, the observable system architecture, and the mass, radius, and orbital period of the innermost detected planet.\\
Building on this study, in \cite{Davoult2025} we trained a Random Forest model on three synthetic populations of planetary systems from the Bern model to identify and classify systems most likely to host an ELP. 
Random Forest is an ensemble learning method composed of multiple decision trees. Each tree (or local classifier) is trained on a random subset of the data, and the global classifier makes decisions based on majority voting among local classifiers. If most trees classify a system as `True' (hosting an ELP), the final classification is `True'; otherwise, it is `False'. 
The Random Forest model used in this study consists of 500 decision trees. This approach balances variance reduction via ensemble learning while maintaining reasonable training time. Each tree is trained on a minimum of 100 instances, increasing diversity among trees and improving generalization. 
Additionally, tree depth is limited to five, meaning the model can make at most five sequential splits from root to leaf. This constraint reduces complexity, prevents overfitting, and ensures the model focuses on capturing the most general and meaningful trends in the data. \\
This approach demonstrated strong performance in model testing, achieving a precision score of up to 0.99, indicating that 99\% of positively classified instances were true positives. This result confirms that the Random Forest model accurately identifies systems hosting an undetected ELP within the synthetic populations of the Bern model.\\
We then applied this model to 1567 observed planetary systems with at least one known planet to identify the most likely ELP-hosting candidates. Among these, the model classified 44 systems with a voting rate $\geq$ 90\%, suggesting a high probability of hosting an ELP. Further details and a complete list of identified systems are available in \cite{Davoult2025}.

%% file: svmult/author/Valentin/main_valentin.tex
Since the beginning of space exploration, the identification and morphometric characterization of atmospheric and surface features on bodies across the solar system has largely relied on human-driven, manual review of orbital image data. Faced with exponentially increasing data downlink rates and image archives with millions of individual images, the human-driven review of image data becomes increasingly tedious and ineffective, effectively reducing the scientific return of current and upcoming missions. Deep learning-driven search routines are a powerful tool to address the limitations of manual image review, representing the only viable way of enabling the processing and analysis of the vast data to be returned by future exploration missions. Since 2018, a range of studies has been exploring the application of deep learning tools to planetary surface and atmospheric mapping, underlining its significant potential.\\
The authors of \cite{Bickel2018,Bickel2020a} demonstrated how off-the-shelf supervised CNN-driven object detection architectures (specifically RetinaNet – TensorFlow and YOLO – PyTorch) can be re-purposed and re-trained to identify \emph{rockfalls} in orbital images of the Moon and Mars, i.e., small geologic expressions of weathering, erosion, and planetary surface evolution.
\cite{Bickel2020b} deployed these CNNs at scale, scanning the entire lunar and martian surfaces (processing hundreds of thousands of orbital images) and identifying more than 130~000 rockfalls, marking the first application of CNNs to create a global-scale catalog of geologic features on the Moon and Mars (or any planetary body). Using the spatial distribution of rockfalls, \cite{Bickel2020b,Bickel2021a,Bickel2024a} inferred the geologic processes driving rockfall occurrence on the Moon and Mars, most notably meteoroid impacts, insolation, and quakes.\\
The same authors developed follow-up CNN detectors to create global catalogs of other geologic expressions of surface evolution on the Moon, specifically granular flows \citep[i.e, small landslides; ][]{Bickel2022} and fractured boulders \citep{Ruesch2023, Ruesch2024}. These studies demonstrated the decisive role of host terrain age and surface composition in controlling the nature and magnitude of planetary surface evolution. Currently, the authors are supporting an effort to use CNNs to catalog every single boulder on the lunar surface \citep{Aussel2024}.\\
On Mars, \cite{Mills2024} and \cite{Bickel2024b} developed CNN detectors to identify geomorphic, geologic, and mineralogic expressions of past (near-)surface water on Mars, specifically pitted cones and chloride salt deposits. These studies revealed that pitted cone formation on Mars is most likely driven by mud rather than igneous volcanism and uncovered new evidence of transient bodies of near-surface brines that were widespread in Mars’ northern hemisphere, with significant implications for our understanding of Mars’ past climate. Additionally, \cite{Wagstaff2022,Bickel2025a,Charalambous2025} mapped hundreds of new impact events and correlated their spatiotemporal occurrence with seismic data from NASA’s InSight mission. These studies identified the largest impacts found near InSight, which were missed by earlier year-long manual surveys, increased the number of potential seismic–impact event matches by a factor of $\sim$8, and revealed that Mars’ present-day impact rate is approximately twice as high as previously estimated. Further, \cite{Conway2025} systematically catalogued atmospheric phenomena such as dust devils in several decades’ worth of orbital image data, offering unprecedented insights into the dynamics of the lowermost martian atmosphere and the sourcing and global distribution of atmospheric dust.\\
On Mercury, \cite{Bickel2025b} employed CNN-driven detectors to create the first consistent, global reference catalog of Mercury’s enigmatic `hollow' features.
This study, along with \cite{Deutsch2025}, provided several independent lines of evidence for the processes that form and shape hollows over time: subsurface volatile exposure through micrometeoroid bombardment and insolation-driven outgassing of subsurface volatiles.\\
The author of \cite{Bickel2021b,Bickel2021c} experimented with and demonstrated the significant potential of multi-domain learning---leveraging data from multiple planetary bodies---to enable the re-utilization of existing labels and reduce the need for the tedious creation of new batches of training and testing labels when studying similar features across different planetary bodies. In addition, \cite{Lesnikowski2024} demonstrated the value of unsupervised detection methods, such as VAEs, which utilize all available data in a dataset without requiring training labels. 

%% file: svmult/author/Caroline/main_caroline.tex
\begin{figure}
    \centering
    \includegraphics[width=\linewidth]{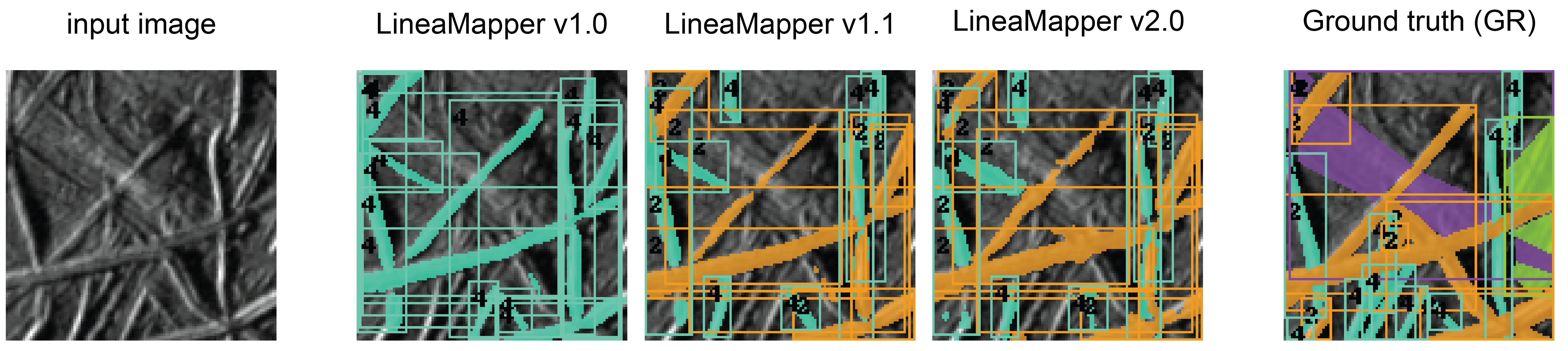}
    \caption{An example of a 112x112 pixel input image of Europa's surface at a regional scale of 215~m/px. The input image shows a dense network of linear surface features. The four colours represent four different categories. The output of LineaMapper v1.0 \citep{Haslebacher2024a} served as a guideline for generating more training data to train v1.1 (Mask R-CNN) and v2.0 (SAM) \citep{HaslebacherPSJ2025}. The ground truth is closest to the output of LineaMapper v1.1, although not all of the $>$15 features were identified. Figure adapted from \cite{HaslebacherPSJ2025}.}
    \label{fig:Haslebacher}
\end{figure}

The automated detection and segmentation of \emph{linear} surface features on planetary bodies can pose different challenges than previously mentioned objects (e.g. boulders, pitted cones, dust devils). First, a linear feature can encompass a large bounding box in which only few pixels belong to the linear feature (imagine it is diagonal to the image borders). Second, if the linear feature is only a few pixels ($\leq$ 5 px) in width, a small inaccuracy of the model segmentation can lead to a very small intersection-over-union (IoU; the IoU denotes the number of pixels in the intersection divided by the number of pixels of the union of predicted mask (or box) and ground truth mask (or box)).
In the particular case of linear surface features on Jupiter's icy moon Europa (lineaments), an intricate network of lineaments at a regional scale (200 m/px) poses a challenging problem to disentangle. In some geologic areas, the ridged plains, the network of lineaments is so dense that one image displaying an area of ca. 580~km$^2$ can contain a large number of lineament instances (see Fig. \ref{fig:Haslebacher}). 

Hitherto, mapping of lineaments was a manual task \citep[e.g., ][]{Figueredo2004, Sarid2004, Kattenhorn2002, Patterson2006, Rhoden2013, Collins2022}. This limits the amount of data that can be analysed and therefore limits the global scale of the drawn conclusions of a given lineament study. Partially, the extent is, as of today, limited by the available imagery from previous space missions (Galileo and Voyager). However, in the next decade, approximately 500 times more imaging data will be available for a global lineament analysis \citep{TurtleSSR, Daubar2024, Haslebacher2024a}. Therefore, preparation for an automated analysis can help to gain near-instantaneous insights and potentially adapt mission planning. \cite{Haslebacher2024a} introduces a supervised deep learning network based on the reliable Mask R-CNN architecture \citep{He2018} that demonstrates the ability of deep learning models to tackle lineament analysis on Europa. Eventually, the automated mapping of lineaments can inform formation mechanisms of geologic units on a global scale, as shown in \cite{HaslebacherPSJ2025} for a selected semi-automatically mapped area.\\
\cite{HaslebacherPSJ2025} also concluded that for this specific task, the performance of a fine-tuned deep learning model is dominated by the dataset, not the architecture, agreeing with observations made by \cite{Bickel2021b,Bickel2021c}. 
\cite{Haslebacher2025MapsBook} applies the model of \cite{Haslebacher2024a} without fine-tuning to other planetary bodies (Venus, Ganymede, Enceladus) to showcase the usability and transferability of linear surface feature detection on planetary bodies.\\
\cite{HaslebacherAzi} introduces an automated extraction algorithm of a linear feature's direction, and length, width, number of segments (used in \cite{HaslebacherPSJ2025}). To conclude, future planetary data from upcoming missions request a more efficient way of extracting information from the data, without eliminating the human from the loop. 

%% file: svmult/author/Emily/main_emily.tex
Spectroscopic observations of substellar companions are essential for unveiling the structure and composition of exoplanetary atmospheres \citep[e.g.,][]{line2016no, brogi2019retrieving}. Such insights are pivotal in understanding the formation history of exoplanets \citep[e.g.,][]{nowak2020peering, molliere2022}. In this regard, cross-correlation spectroscopy (CCS) has emerged as a well-established technique for molecular detection and atmospheric characterisation.\\
This method involves cross-correlating a molecular template or full atmospheric model with an observed planetary spectrum across a range of radial velocities (RVs), with a detection indicated by a statistically significant correlation peak at the planet’s RV. 
CCS is recognised as a well established method for detecting molecular species and isotopologues within atmospheres \citep[e.g.,][]{konopacky2013detection, de2013detection, molliere2019detecting}. In this case, CCS is particularly useful when the planet’s continuum is lost during data reduction \citep[e.g.,][]{konopacky2013detection, cugno2021molecular, malin2023simulated}, for example in absence of angular differential imaging \citep[ADI,][]{marois2006angular}.\\
While CCS and molecular mapping have demonstrated robust performance in identifying molecular species, the statistical significance of these detections often relies on the signal-to-noise (S/N) ratio of the cross-correlation peak. However, this metric typically assumes Gaussian, independent, and identically distributed (i.i.d.) noise, which is often not the case due to systematic and random effects from instrumentation, observation conditions, and stellar or molecular residuals \citep{malin2023simulated}. Such violations of noise assumptions can lead to unreliable detection confidence when using traditional S/N metrics \citep{Bonse_2023}. In addition, it is not possible to automatise the S/N scoring  of a high number of spectra from multiple datasets or planets, as the application of this S/N metric relies on prior knowledge of the expected location of the cross-correlation peak.\\
Such challenges were recently discussed and addressed by \cite{garvin2024machine} using ML techniques in the spectral dimension and \cite{nath2024} in the spatio-temporal regime. They show together that learning relevant features from cross-correlated spectral data can enhance detection capabilities compared to the use of traditional S/N-based metrics. In particular, \cite{garvin2024machine} introduced the ML for cross-correlation spectroscopy (MLCCS) approach, integrating ML with CCS to enhance the sensitivity to faint molecular signals. Using one dimensional CNNs, this method analyses the full RV dimension to detect deterministic patterns from individual molecular features, especially in non-Gaussian and non-i.i.d. noise environments. In addition, the method was proven to be invariant to shifts in RV and is this applicable to multiple planets and spectral instances at a time.
The MLCCS method offers enhanced detection sensitivity and reliability for detection of individual molecules, and conspicuity in molecular maps. This method is expected to be particularly useful with IFS instruments like JWST/MIRI \citep{patapis2021direct, malin2023simulated}, VLT/ERIS\citep{hayoz2025high}, and the upcoming ELT/HARMONI \citep[e.g.,][]{houlle2021direct, bidot2024exoplanets, vaughan2024behind} but its one-dimensional framework is designed to be adaptable to slit spectroscopy \citep[e.g., with Keck/OSIRIS and CRIRES+][]{de2014identifying, dit2018molecule, boldt2023optimising} as well as individual spectra. This method's agnosticism towards specific atmospheric models and its robustness to complex noise conditions could lead to a paradigm shift in how molecular detections are validated, potentially optimizing telescope time and improving the efficiency of spectroscopic surveys \citep{garvin2024machine}. 


%% file: svmult/author/Salome/main_salome.tex
\begin{figure}[h!]
\sidecaption
\includegraphics[width=11.7cm]{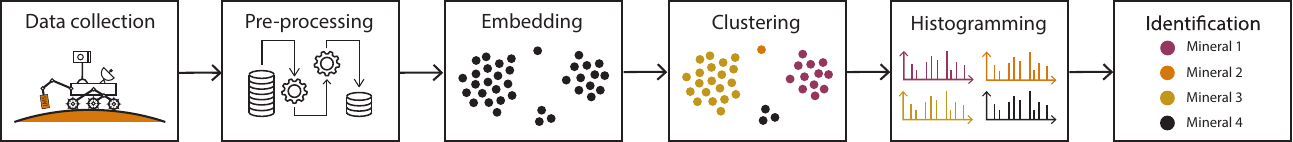}
\caption{Unsupervised analysis pipeline for on-board data clustering. The collected and pre-processed data can be dimensionality reduced and clustered to find the groups of spectra containing information of interest, for example, signatures of certain minerals or trace elements. Reproduced and adapted from \cite{Gruchola2024}.}
\label{fig:Gruchola}
\end{figure}

Raw scientific data are, in most cases, unlabeled, and the generation of labels often involves labor-intensive manual work. 
In cases where labeled data are scarce, difficult to obtain, or impractical to generate---such as in space research---unsupervised ML becomes particularly useful. As unsupervised ML only requires the data themselves as input, without pre-existing labels, it is a powerful tool for identifying hidden patterns and structures within datasets and understanding their distribution in high-dimensional space. The most prominently used unsupervised techniques include dimensionality reduction, density estimation, and clustering, which allow to extract meaningful insights from raw, high-dimensional data.\\
Dimensionality reduction simplifies complex data sets by transforming them into a lower dimensional space while preserving essential information. This transformation can exploit correlations or similarity between features (based on a chosen metric). By reducing the number of features, dimensionality reduction helps visualize high-dimensional data and reduces the computational requirements for subsequent analyses, such as clustering. Clustering groups data points on the basis of common characteristics, allowing the homogeneity of a data set to be examined effectively. While clustering of high-dimensional data is possible, removing or reducing features that carry little or no information generally improves the quality of the clustering analysis, hence, these algorithms often benefit from being applied to dimensionally reduced data.\\ 
Both dimensionality reduction and clustering techniques can be applied to a wide range of data. In mass spectrometry, \cite{Gruchola2024} uses dimensionality reduction and clustering to separate collected mass spectra into distinct groups based on their chemical composition.
These techniques have been applied to data collected from microfossils in ancient geological samples from Earth \citep{Lukmanov2021} using a space-prototype laser ablation and ionization mass spectrometer (LIMS) developed at the University of Bern \citep{Riedo2013}. 
Data preprocessing, including noise reduction and normalization, has been shown to be critical to ensure effective data handling and to reduce the influence of noise on the results \citep{Gruchola2024}. Such applications highlight the potential of unsupervised ML to identify biosignatures and mineralogical features in extraterrestrial materials.
In the context of space exploration, unsupervised ML still plays a minor role compared to supervised approaches. So far, it has mainly been applied to space-acquired data, but in-situ applications are becoming more popular. These include data analysis, anomaly detection, and resource optimization \citep{Labreche2022, Murphy2023,Ruzicka2022}. These techniques can support data prioritization, allowing spacecraft to optimize the selection of high-value or interesting data for transmission to Earth, as visualized in Figure \ref{fig:Gruchola}.

%% file: svmult/author/JoAnn/main_joann.tex
A central question in exoplanet research is what the interiors of exoplanets look like. On the one hand, this is because the detected exoplanet population exhibits a remarkable diversity in mass and radius, with many of the discovered planets not having analogues in the solar system. On the other, exoplanet interiors are shaped by the properties of the protoplanetary discs in which they formed, as well as their formation locations, orbital migration, and evolution histories. As a result, studying the composition of exoplanets can give us insights into planet formation and evolution processes, although the exact connection remains complex. One of the main challenges when trying to study exoplanet interiors is the inherent degeneracy in determining a planet's internal composition from observations. Generally, our knowledge of the properties of exoplanets from observations is sparse; in most cases, we will be limited to knowing a planet's radius and mass, giving us an estimate of its main density, and its orbital period, informing us of its temperature. While this gives us a general idea of what the planet could look like, there is a very large number of interior compositions that would be compatible with these observed quantities \citep[e.g.][]{Seager+2007, Rogers+Seager2010a}. Without additional observations, we cannot break this degeneracy, no matter how precisely the mass and radius values in question are known. We can, however, try to reduce it by excluding structures that are unphysical and using any additional information available, such as the composition of the planet's host star \citep[e.g.][]{Dorn2015}.\\ 
To this end, a wide variety of planetary structure models have been developed over the last two decades to describe mass--radius relations of various types of planets \citep[e.g.,][]{Valencia+2006,Seager+2007,Sotin+2007,Fortney+2007,Dorn2015,Dorn2017,Zeng+2016,Brugger+2017,Acuna+2021,Aguichine+2021,Dorn+Lichtenberg2021,Haldemann+2024}. Traditionally, Bayesian inference algorithms, such as Markov Chain Monte Carlo (MCMC) methods, have been used in combination with these forward models to infer the internal structures of observed exoplanets \citep[e.g.,][]{Dorn2015,Dorn2017, Acuna+2021, Haldemann+2024}. However, MCMC-based inference schemes require significant computational resources, often taking hours or days to converge.\\ 
More recently, also ML approaches have been used to infer the internal structure of observed exoplanets. \cite{Baumeister+2020} provided proof-of-concept for using a mixture density network to predict the parameter space of compatible interior structures, with their newer work presenting a more physically elaborate model \citep{Baumeister+Tosi2023}. Similarly, \cite{Haldemann+2023} train a conditional invertible neural network with the same purpose. The main advantage of these methods over traditional inference schemes is that the inference can be completed within seconds \citep{Baumeister+Tosi2023} or minutes \citep{Haldemann+2023}. While these previous works use ML to replace the full inverse scheme, the open-source plaNETic framework\footnote{\url{https://github.com/joannegger/plaNETic}} \citep{Egger+2024} instead replaces the forward model with a neural network (NN).\\
plaNETic is based on the planetary structure model of the BICEPS code \citep{Haldemann+2024}, which was used generate a large database of 15~million planetary structure models as training data for a conventional feed-forward NN with multiple hidden layers. Once trained, this NN is then used as a computationally efficient surrogate model for the planetary structure model in a full-grid accept-reject sampling scheme. The trained NN can predict planetary radii up to 50~000 times faster than direct numerical solutions of the planetary structure equations. A proof-of-concept of the method was introduced and applied to a system of observed planets for the first time in \cite{Leleu2021}, while the updated version of the framework, featuring the full physical complexity of the BICEPS planetary structure model, was presented in \cite{Egger+2024}, along with its first application to an observed planetary system. Both versions of the framework together have been applied to infer the interiors of exoplanets in various systems, presented in more than 30 publications \citep[e.g.,][]{Bonfanti+2023, Patel+2023, Luque+2023, Bonfanti+2024_toi396,Stalport+2025,Zingales+2025}. Some of these results are summarised in more detail in the chapter by Leleu et al. from the same collection. \\
Compared to the hours or days that are needed for the inference process with MCMC, with plaNETic it takes less than 30 minutes to infer the interior structure of an observed planet. 

%% file: svmult/author/Yann/mass_prediction.tex
Computing the formation of planetary systems in the Bern model relies on different modules (see chapter by Kessler et al., this collection): the structure and evolution of the protoplanetary disk, the accretion of solids and gas, or the planet-planet interaction among others. One of these modules is the determination of the gas accretion of forming planets. Gas accretion result from two effects: 1- a planet accreting solids (planetesimals and pebbles) shows an increase gravitational potential well, which tends to contract the planetary envelope, while 2- the outer edge of the planet (which depends on the planetary Roche and/or Bondi radius) increases with the planetary mass. Computing gas accretion of a given planet therefore requires  determining the total planetary mass such as the outer planetary radius matches its outer radius (Roche radius or Bondi radius). This boils down to computing, as a function of several parameters, the total planetary mass that fulfils all the boundary conditions (core mass, core luminosity,  disk temperature and pressure at the planet location,  planet location itself). Such computation is done by solving the differential equations describing the internal structure of a planet (similar to the ones used in the previous section - but with different boundary conditions) using an iterative scheme. The planetary mass is the value that allows matching all boundary conditions\footnote{Note that there are in general more than one values that allow matching all conditions - see \cite{Venturini2015}}. The other approach often used in the litterature is to use scaling lows (e.g. \cite{Bitsch_etal_2015}), but these laws turn out to be rather imprecise.\\
When computing the formation of a planetary system, the internal structure of planetary systems is computed millions of times (for all planets at each timestep), with boundary conditions that are not extremely different from one case to another. In order to accelerate this computation, members of the NCCR PlanetS have trained a DNN that predicts the mass of a planet as a function of the accretion rate of solids, the core mass, and the disk properties at the location of the planet. The approach followed is similar to the ones presented in the previous section, with 1- the computation of a large database of models, and 2- the training of a DNN (see \cite{Alibert_Venturini_2019}). Using this approach, the authors demonstrated very good performances compared to scaling laws, speeding up the internal structure calculations by orders of magnitude \citep[see e.g., ][]{Venturini_etal_2020}.
